\documentclass[amssymb,aps,twocolumn,showpacs, superscriptaddress]{revtex4}
\usepackage[]{graphicx}
\usepackage[]{amsmath}
\usepackage{hyperref}

\def\sket#1{| #1 \rangle\hspace*{-2pt}\rangle}
\def\sbra#1{\langle\hspace*{-2pt}\langle #1 |}
\def\skb#1#2{| #1 \rangle\hspace*{-2pt}\rangle\!\langle\hspace*{-2pt}\langle #2 |}
\def\sbracket#1#2{\langle\hspace*{-2pt}\langle #1| #2\rangle\hspace*{-2pt}\rangle\!  }

\def\cB{\mathcal{B}}
\def\cC{\mathcal{C}}

\def\cH{\mathcal{H}}
\def\cJ{\mathcal{J}}

\def\cL{\mathcal{L}}
\def\cM{\mathcal{M}}

\def\cO{\mathcal{O}}

\def\cS{\mathcal{S}}
\def\cT{\mathcal{T}}

\def\Tr{\mathrm{Tr}}

\def\Re{\mathbb{R}}

\def\eq#1{Eq.~\eqref{eq:#1}}

\begin{document}

\title{Lieb-Robinson Bound and Locality for General Markovian Quantum Dynamics}
\author{David Poulin}
\email{David.Poulin@USherbrooke.ca}
\affiliation{D\'epartement de Physique, Universit\'e de Sherbrooke, Qu\'ebec, Canada}

\date{\today}

\begin{abstract}
The Lieb-Robinson bound shows the existence of a maximum speed of signal propagation in discrete quantum mechanical systems with local interactions. This generalizes the concept of relativistic causality beyond field theory, and provides a powerful tool in theoretical condensed matter physics and quantum information science. Here, we extend the scope of this seminal result by considering general Markovian quantum evolution, where we prove that an equivalent bound holds. In addition, we use the generalized bound to demonstrate that correlations in the stationary state of a Markov process decay on a length-scale set by the Lieb-Robinson velocity and the system's relaxation time. 
\end{abstract}

\pacs{05.30.-d,03.67.-a}

\maketitle

In relativistic quantum field theory, the actions of an observer can only influence his future light-cone. This can be seen as a consequence of the fact that interactions are {\em covariant} and {\em local}, i.e., they couple the field at a given point only to the field at points located infinitesimally close to it. The situation is, at first glance, quite different for discrete quantum mechanical systems with local interactions, such as spin lattices with nearest neighbor couplings. There, it is in principle possible to send information between any two connected regions in an arbitrarily short time, despite the fact that interactions are local. However, a result first derived by Lieb and Robinson \cite{LR72a}, and improved in Refs. \cite{H04b,H04c,HK06a,NOS06a}, demonstrates the existence of an effective light-cone such that the amount of information signaled beyond it decays exponentially. 

This result, known as the Lieb-Robinson bound, is derived under the assumption of unitary evolution, i.e., when the dynamics is governed by Schr\"odinger's equation. A more general form of quantum evolution is given by Markovian dynamical semigroup equations, that are the natural generalizations of stochastic processes to the quantum setting. These are needed, for instance, to describe systems with dissipation or decoherence, and include unitary evolution and classical stochastic evolution as special cases. 

In this Letter, we extend the Lieb-Robinson bound to general local Markovian dynamics. Moreover, we demonstrate that the correlations displayed in the stationary state of a Markov process decay exponentially beyond a length-scale set by the Lieb-Robinson velocity and the system's relaxation time, which can be related to the gap of the semi-group generator. This also generalizes the results of Hastings \cite{H04c} established in the setting of {\em classical} Markovian dynamics.

There are several motivations to study the existence of an effective light-cone under general quantum dynamics. First, experimental systems are always subject to some amount of dissipation and decoherence, so understanding the origin of causality under these conditions is important. Second, the Lieb-Robinson bound has proven to be a powerful tool to characterize the structure of ground states of gapped Hamiltonians. For instance, Hastings and collaborators have used this bound to rigorously prove the stability of topological order \cite{BHM10a}, the existence of PEPS representation of ground states \cite{Has06a}, the exponential decay of correlations \cite{H04c}, and to generalize the Lieb-Schultz-Mattis theorem to higher dimensions \cite{H04b}. One might naturally expect similar characterizations of thermal states to emerge from the current work; in fact, we provide a first step by proving clustering of correlations for the fixed points of gapped Markov processes. 

Thirdly, the Lieb-Robinson bound for unitary processes is an important tool in quantum complexity theory, see e.g. \cite{H09a}. It was recently demonstrated that dissipation is a universal resource for quantum computation \cite{VWC09a}; our results complement this finding in a natural way. One major open question in this field, related to the quantum PCP conjecture, is to identify the complexity of finding the ground state energy density of a local Hamiltonian within constant accuracy. A problem that is at least as hard can be formulated in terms of thermal states, so our result could shed new light on this open question. 

Lastly, the existence of a fundamental minimal length scale, the Planck length, suggests that physics might be fundamentally discrete; many approaches to quantum gravity have this discreteness built in (e.g. \cite{L98a} and references therein). The black-hole evaporation problem also suggests that quantum mechanics could be fundamentally non-unitary, with unitary dynamics emerging as a low energy approximation \cite{Haw82a}. Our result provides a mechanism for emergent causality in  such fundamentally {\em discrete} and {\em non-unitary} theories. 


\medskip
\noindent{\em Lieb-Robinson bound}---We consider the setting where particles are located over a set of vertices $\Lambda$. The particle at location $x\in \Lambda$ has Hilbert space $\cH_x$, so the entire Hilbert space is $\bigotimes_{x\in \Lambda} \cH_x$. For any subset of vertices $X\subset \Lambda$, we write $\cH_X = \bigotimes _{x\in X} \cH_x$. We make no distinction between operators $O_X \in \cB(\cH_X)$ on $X$ and their natural embedding in $\cB(\cH)$. A metric $d(x,y)$ is defined between particles locations. A good example to keep in mind are spins located at the vertices of a regular $D$-dimensional lattice, where $d(x,y)$ is the usual graph distance. The Hamiltonian is given by a sum of terms $\sum_{X\subset \Lambda} H_X$ where $H_X = 0$ for all $X$ of diameter greater than some constant $d^*$. Recall that the diameter of a subset of vertices is given by the largest distance between any pair of vertices inside it. Thus, $H$ is ``local" or ``short-ranged" in the usual sense. 

To motivate the statement of the Lieb-Robinson bound, imagine that one observer, Alice, has access to some particles $A \subset \Lambda$ and wants to signal to a second observer, Bob, who has access to $B\subset \Lambda$. The system is initially in the state $\rho \in \cB(\cH)$. To send the signal ``0", Alice does nothing, while if she wants so signal ``1", she applies a transformation to the particles in her possession, mapping the state $\rho$ to $ \rho' =  \rho + i\epsilon  [O_A,\rho]$ where $O_A \in \cB(\cH_A)$. To read the signal after some time $t$, Bob must perform a measurement on region $B$ that discriminates between the state $\rho(t)$ and $\rho'(t)$, where time evolution is governed by Schr\"odinger's equation $\dot\rho(t) = -i[H,\rho(t)]$. If he makes a measurement described by the operator $O_B \in \cB(\cH_B)$, the probability that he distinguishes the two signals is $|\Tr\{O_B [\rho(t) - \rho'(t)]\}| = \epsilon| \Tr\{\rho [O_B(t),O_A]\}| \leq \epsilon \| [O_B(t),O_A]\|$, where time evolution in the Heisenberg picture is governed by $\dot O(t) = i[H,O(t)]$. 

The Lieb-Robinson bound shows that 
\begin{equation}
\| [O_B(t),O_A]\| \leq c V \|O_A\|\|O_B\| \exp\left\{-\frac{d_{AB}-vt}\xi\right\}
\label{eq:LR}
\end{equation}
where $d_{AB}$ is the distance between the regions $A$ and $B$, $V = \min\{|A|,|B|\}$ is the volume of the smallest of the two regions, and $c$, $v$, and $\xi > 0$ are constants that depend only on the microscopic details of the model: the interaction strength $\max_{X\subset \Lambda} \|H_X\|$, the radius of interactions $d^*$, and the maximal degree of the vertices. Thus, signals can only propagate at a finite velocity $v$, defining an effective light-cone. Outside this cone, the probability of detecting a signal falls off exponentially. 


\medskip
\noindent{\em Markov dynamical semigroup equations}---We now generalize the setting by considering a broader class of evolution equations. Lindblad has shown \cite{Lin76a} that the most general differential equation for $\dot\rho$ that 1) is linear, 2) is local in time (Markovian), 3) preserves positivity, and 4) preserves the trace must have the form
\begin{equation}
\dot\rho = -i[H,\rho] + \sum_a L_a \rho L_a^\dagger -\frac 12 \left( L_a^\dagger L_a \rho + \rho L_a^\dagger L_a\right)
\end{equation}
where $H$ is a Hamiltonian and $L_a$ are any operators. In general, the Hamiltonian and the operators $L_a$ can be time-dependent; our result holds in that case as well but we consider time-independent generators for simplicity. In the Heisenberg picture, this equation gives 
\begin{align}
\dot O &= i[H,O] + \sum_a L_a^\dagger O L_a -\frac 12 \left( L_a^\dagger L_a O + O L_a^\dagger L_a\right)
\label{eq:Lindblad} \\
&=: \cL [O].
\end{align}

It is convenient to adopt a super-operator notation, viewing $\cB(\cH)$ as a vector space. For $O\in \cB(\cH)$, we use the notation $\sket O$ and denote the Hilbert-Schmidt inner product $\sbracket O{O'} = \Tr\{O^\dagger O'\}$. Then, we can express \eq{Lindblad} as $\sket{\dot O} = \cL \sket O$, and the formal solution is $\sket{O(t)} = e^{\cL t} \sket{O(0)}$. We will often switch between the two notations. 

Like in the original setting, we are interested in the case where the time-evolution generator is given by the sum of local pieces, $\cL = \sum_{X\subset \Lambda} \cL_X$ with $\cL_X = 0$ for all $X$ of diameter greater than some constant $d^*$. Each term $\cL_X \in \cB(\cB(\cH_X))$ in that sum has the Lindblad form \eq{Lindblad} with $H_X, L_{X,a} \in \cB(\cH_X)$.
We make the assumption \footnote{The norm we use on super-kets $\sket O$ is the corresponding operator norm $\|\sket O\| = \|O\| \neq \sqrt{\sbracket OO}$. The norm we use for super-operator is the induced super-operator norm $\|\cL\| = \max\left\{\frac{\|\cL \sket O\|}{\|\sket O\|}\right\}$.} throughout that $\|\cL_X\| \leq 1$, which is equivalent to fixing the time units. 


\medskip
\noindent{\em Lieb-Robinson bound for Markov processes}---We now come to the main result, which is a bound on $[O_B(t),O_A]$ where the dynamics of $O_B$ is governed by a local Markov process as described above. Our proof is inspired by that of \cite{HK06a,NOS06a}. The main complication comes from the fact that quantum dynamical semigroup equations do not obey Leibniz rule $\frac \partial{\partial t} (O_AO_B) \neq \frac {\partial O_A}{\partial t} O_B + O_A  \frac {\partial O_B}{\partial t}$, and that backward time evolution is norm-increasing. Here, we will only present parts of the proof that are distinct from the unitary case, other details can be found in  \cite{HK06a,NOS06a}.

We are interested in the quantity $f(t) = [O_B(t),O_A]$ that we can expressed as $f(t) = \cC_A e^{\cL t} \sket{O_B}$, where $\cC_A$ is the super-operator defined by the action  $\cC_A\sket Q := \sket{[Q,O_A]}$.  We can write a differential equation for $f(t)$
\begin{align}
\dot f(t) &= \cC_A \cL e^{\cL t} \sket {O_B} \\
& = \cL_{\bar A} \cC_A e^{\cL t} \sket {O_B} + \cC_A \cL_{\cap A} e^{\cL t} \sket {O_B} \\
& = \cL_{\bar A} f(t) + \cC_A \cL_{\cap A} e^{\cL t} \sket {O_B}
\end{align}
where we have broken the Lindblad super-operator in two parts, $\cL_{\cap A} = \sum_{X: X\cap A \neq 0} \cL_X$ and $\cL_{\bar A} = \cL - \cL_{\cap A}$, and we used the fact that $[\cC_A,\cL_{\bar A}] = 0$.

It can easily be verified by differentiating that the solution to this differential equation is
\begin{equation}
f(t) = e^{\cL_{\bar A} t}f(0)+\int_0^t e^{\cL_{\bar A} (t-s)} \cC_A \cL_{\cap A} e^{\cL s} \sket {O_B} ds.
\end{equation}
Because $e^{\cL_{\bar A} t}$ is norm-contracting for $t\geq 0$, it follows that
\begin{equation}
\| f(t)\| \leq \|f(0)\| + \|\cC_A\| \int_0^t \| \cL_{\cap A}e^{\cL s} \sket {O_B} \| ds.
\label{bound_f}
\end{equation}
We can now recurse. Define the quantity
\begin{equation}
M_O(X,t) = \sup_{\cT \in \mathbb{L}_X} \frac{\|\cT e^{\cL t} \sket O\|}{\|\cT\|}
\end{equation}
where $\mathbb{L}_X$ is the set of super-operators $\cT$ of the form \eq{Lindblad} with $L_a$ and $H$ $\in \cB(\cH_X)$. With this definition, it follows that for any $X\cap Y = 0$, 1) elements of $\mathbb{L}_X$ and $\mathbb{L}_Y$ commute, and 2) elements of  $\mathbb{L}_X$ annihilate $\cB(\cH_Y)$.

Repeating the steps leading to Eq.~\ref{bound_f}, we have 
\begin{equation*}
M_{O_B}(X,t) \leq M_{O_B}(X,0) + \sum_{Y: Y\cap X\neq 0} \int_0^t M_{O_B}(Y,s) ds.
\end{equation*}
From here, the arguments of \cite{HK06a,NOS06a} can be used to show that $[O_B(t),O_A]$ is bounded by \eq{LR}. Note that, using the techniques of \cite{PHKM09a}, it should be possible to generalize this bound to the case that the Lindblad super-operator is the sum of local {\em unbounded} terms with bounded commutators. 


\medskip
\noindent{\em Convergence rate}---Before we examine the correlations generated by quantum Markov processes, a few words about their asymptotic properties are in order. The Lindblad super-operator can be written in its Jordan normal form  $\cL = \cS \cJ \cS^{-1}$ where $\cJ = \bigoplus_{j\geq 0} \cJ_{d_j}(\lambda_j)$ is the Jordan matrix and we choose $\Re(\lambda_0) \geq \Re(\lambda_1)\geq \ldots \geq \Re(\lambda_k)$. Trace preservation implies that $\Re(\lambda_0) = 0$. If the Markov process has a unique stationary state $\pi$  such that $\sbra\pi e^{\cL t} = \sbra\pi$, then the first Jordan block is one-dimensional, $d_0 = 1$, and the gap $\Delta = -\Re(\lambda_1)$ is strictly positive \cite{BNPV10a,TD00a}.  Asymptotically, the system converges to this unique stationary state $\lim_{t\rightarrow \infty} e^{\cL t} = \cS {\rm{diag}}(1,0,\ldots,0)\cS^{-1} =  \skb{I}{\pi}$. The gap $\Delta$ governs the rate of convergence to equilibrium:
\begin{equation}
e^{\cL t} - \skb I\pi = \cS \bigoplus_{j>0} e^{\lambda_j t} \cM_{d_j} \cS^{-1}
\end{equation} 
where $\cM_d$ is the $d\times d$ matrix with 1's on the diagonal and $\frac 1{k!}$ on its $k$th upper diagonal. Because $\|\cM_d\| \leq e$, we conclude that $\|e^{\cL t} - \skb I\pi\| \leq \|\cS\| ^2 e^{-\Delta t +1}$ (note that we can always choose $\cS$ such that $\|\cS\|=\|\cS^{-1}\|$, which we assume henceforth). Thus, the inverse gap of $\cL$ sets the relaxation rate, but the pre-factor $\|\cS\|^2$ can scale with the system size in the case that $\cS$ is ill-conditioned. Finding conditions that make the conditioning number of $\cS$ constant for local $\cL$ is an interesting question that we leave open.


\medskip
\noindent{\em Clustering of correlations}---We now demonstrate that, when the system has a relaxation time $\tau$ that is independent of the system size, the fixed state $\pi$ exhibits clustering of correlations in the sense that $\langle O_AO_B\rangle \approx \langle O_A\rangle \langle O_B\rangle $ for operators supported on regions $d_{AB} \gg v\tau$ apart. This will occur for instance when $\cL$ is gapped and $\|\cS\|$ constant.

Starting in any initial state $\rho$, the system reaches the stationary state $\pi$ in time $t\approx \Delta^{-1}$, so 
\begin{equation}
\Tr\{[\pi-\rho(t)] O_AO_B\} \leq \|O_A\|\|O_B\| \|\cS\|^2e^{-\Delta t}.
\label{eq:relax}
\end{equation} 
In particular, we can choose $\rho$ to be a product state, i.e one without any correlations at all. To gain some intuition, we shift to the Heisenberg picture, where we know that both $O_A(t)$ and $O_B(t)$ grow in space at a speed $v$. Hence, provided that regions $A$ and $B$ are separated by $d_{AB} \gtrsim v\Delta^{-1}$, the operators $O_A(t)$ and $O_B(t)$ will still be supported on disjoint regions by the time the system equilibrates, so $\Tr\{\rho O_A(t)O_B(t)\} \approx \Tr\{\rho O_A(t)\}\Tr\{\rho O_B(t)\}$ for any product state $\rho$. 

The problem with this intuitive argument is that, due to the failure of Leibniz' rule, the operator $(O_AO_B)(t)$---solution to the differential equation $\frac\partial{\partial t} X(t) = \cL[X]$ with $X(0) = O_AO_B$---is not equal to $O_A(t)O_B(t)$. The crucial observation however is that Leibniz' rule holds for any operators $O_A(t)$ and $O_B(t)$ contained on regions separated by at least the interaction range $d^*$. In that case, we have $\cL[O_A(t)O_B(t)] = \cL[O_A(t)]O_B(t) + O_A(t)\cL[O_B(t)]$. Our generalized Lieb-Robinson bound shows that $O_A(t)$ and $O_B(t)$ remain inside their respective light-cones, save for an exponentially decaying tail, so the approximation $(O_AO_B)(t) \approx O_A(t)O_B(t)$ is valid for short times. 

\begin{figure}[tb]
\includegraphics[scale=0.85]{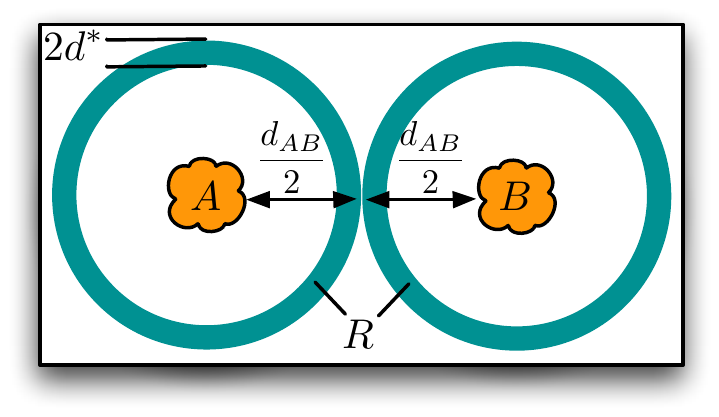}
\vspace*{-0.5cm}
\caption{The region $R$ is the union of two membranes surrounding regions $A$ and $B$ respectively. If we turn off $\cL$ on $R$, the regions $A$ and $B$ are dynamically decoupled.}
\label{fig:R}
\end{figure}

Rigorously, consider the region $R$ that is the union of two membranes of thickness $2d^*$, the first surrounding region $A$ at a distance $d_{AB}/2$ from $A$, and the second surrounding region $B$ in a similar manner, see Figure \ref{fig:R}. We write the Lindblad super-operator as the sum of two terms, the part supported on $R$, $\cL_R = \sum_{X\subset R} \cL_X$, and the rest $\cL_{\cap \bar R} = \cL - \cL_R$. Define $\cL(\eta) = \cL_{\cap \bar R} + \eta \cL_R$, such that $\cL(1) = \cL$ and $\cL(0)$ is the Lindblad super-operator obtained by turning off all terms supported on $R$. It is clear that the evolution generated by $\cL(0)$ cannot correlate regions $A$ and $B$ because any operators on those regions remain confined inside the regions enclosed by the membrane. Thus, for any initial state $\rho$ in which the two regions enclosed by the membranes are not correlated---such as a product state---we have $\sbra\rho e^{\cL(0)t}\sket{O_AO_B} = \sbra\rho e^{\cL(0)t}\sket{O_A}\sbra\rho e^{\cL(0)t}\sket{O_B}$ for all $O_A \in \cB(\cH_A)$, $O_B\in \cB(\cH_B)$, and all $t$. 

Using the integral representation
\begin{equation}
e^{\cL t} = e^{\cL(0)t}+\int_0^1 \int_0^t e^{\cL(\eta)(t-\beta)} \cL_R e^{\cL(\eta)\beta} d\beta d\eta,
\end{equation}
we can express the time-$t$ correlation $\sbra\rho e^{\cL t} \sket{O_AO_B}$ as the sum of two terms. The first $\sbra\rho e^{\cL(0) t} \sket{O_AO_B}$ displays no correlations as explained above. The second can be bounded using the generalized Lieb-Robinson bound:
\begin{align}
&\left|\int_0^1 \int_0^t \sbra\rho e^{\cL(\eta)(t-\beta)} \cL_R e^{\cL(\eta)\beta}\sket{O_AO_B} d\beta d\eta\right| \label{eq:AB1} \\
\leq & \int_0^t \int_0^1  \left\| \cL_R e^{\cL(\eta)\beta}\sket{O_AO_B}\right\| d\eta d\beta  \label{eq:AB2} \\
\leq &  \int_0^t c V \|\cL_R\| \|O_A\| \|O_B\| \exp\left\{-\frac{d_{AB}-2v\beta}{2\xi}\right\} d\beta  \label{eq:AB3} \\
\leq &c V \|\cL_R\| \|O_A\| \|O_B\| \frac\xi v \exp\left\{-\frac{d_{AB}-2vt}{2\xi}\right\}  \label{eq:AB4}
\end{align}
where $V = \min\{|A|+|B|,|R|\}$ and the other constants are as in \eq{LR}. For a $D$-dimensional regular lattice, $\|\cL_R\| \leq c d_{AB}^{D-1}$, where $c$ is a constant that depends on the microscopic details of the model. In general, we will find $\|\cL_R\| = {\rm poly}(d_{AB})$. Combining this bound with \eq{relax} yields the desired result 
\begin{align}
&\langle O_AO_B\rangle  - \langle O_A\rangle\langle O_B\rangle \\
&= \|O_A\|\|O_B\| \cO\left(\|\cS\|^2 e^{-\Delta t} + cV \|\cL_R\| e^{-\frac{d_{AB}-2vt}{2\xi}}\right)\\
& \leq \|O_A\|\|O_B\| \cO\left(\left[\frac{\|\cS\|^2}{V \|\cL_R\|}\right]^{\frac \xi\mu} e^{-\frac{d_{AB}}{2\mu}}\right).
\end{align}
with $\mu = v\Delta^{-1}+\xi$. 

We note that some fairly loose bounds have been used in this derivation and a tighter bound may be achievable. In particular, we ignored the fact that some initial states $\rho$ reach equilibrium much more rapidly than others. We could optimize this choice to improve the bound \eq{relax}, subject to the constraint that the two regions enclosed by the membranes be initially uncorrelated. A natural guess would be to choose the tensor product of the marginals of $\pi$ over the three regions delimited by the membrane.This choice could perhaps compensate for an ill-conditioned $\cS$. We note however that there appears to exist some local gapped Markov model  with long-range correlations \cite{Hastings:private}. Thus, the dependence of the correlations on $\|\cS\|$ may be unavoidable. 

\medskip
\noindent{\em Conclusion}---The principle of relativistic causality is a pillar of modern physics. The Lieb-Robinson bound shows that the principle extends beyond relativistic quantum field theory, to the setting of discrete quantum systems with a Hamiltonian that is the sum of local pieces. Here, we have generalized the Lieb-Robinson bound by considering a broader family of dynamical systems, namely local Markovian quantum evolution, that include unitary quantum evolution and classical stochastic evolution as special cases. The proof of the Lieb-Robinson bound under these conditions differs from the original one due to the breakdown of Leibniz' rule and of the group properties of the time-evolution operator. 

We have used our generalized bound to demonstrate that the correlations displayed in the fixed point of a Markov process decay exponentially on a length-scale set by the system's equilibration time and the Lieb-Robinson velocity. While the latter depends only on the microscopic details of the model, the former can in general scale with the system size even when the generator of the Markov process is gapped. Describing conditions under which the equilibration time is set by the gap of the generator remains an interesting open question. 


\medskip
\noindent{\em Acknowledgements}---The idea of generalizing the Lieb-Robinson bound came during discussions with Alioscia Hamma. I thank him and Matt Hastings for useful conversations. This work is partially funded by NSERC and FQRNT.


\end{document}